\documentclass[prl,twocolumn,showpacs,preprintnumbers,amsmath,amssymb]{revtex4}

\usepackage{graphicx}
\usepackage{dcolumn}
\usepackage{bm}

\begin{document}

\newcommand*{\cm}{cm$^{-1}$\,}


\title{In-plane optical spectroscopy study on FeSe epitaxial thin film grown on SrTiO$_3$ substrate}

\author{R. H. Yuan}
\author{W. D. Kong}
\author{L. Yan}
\author{H. Ding}
\author{N. L. Wang}

\affiliation{Beijing National Laboratory for Condensed Matter
Physics, Institute of Physics, Chinese Academy of Sciences,
Beijing 100190, China}


\begin{abstract}

We perform in-plane optical spectroscopy measurement on (00l) FeSe thin-film
grown on SrTiO$_3$ substrate by pulsed laser deposition method.
The study indicates that the low frequency conductivity consists of two Drude 
components, a broad one which takes up most
of the spectral weight and a very narrow one roughly below 100-150 \cm. 
The narrow Drude component locates at so low frequencies that no
such behavior was observed in iron pnictides. 
The overall plasma frequency is found to be smaller than the FeAs based compounds,
suggesting a stronger correlation effect. Similar to iron pnictides, a 
temperature-induced spectral weight transfer is observed for FeSe.
However, the relevant energy scale is lower. Additionally, 
different from a recent ARPES measurement which revealed a spin density wave (SDW) order
at low temperature for FeSe thin films grown on SrTiO$_3$ substrate, 
no signature of the band structure reconstruction arising from the
formation of the SDW order is seen by optical measurement 
in the thick FeSe films.

\end{abstract}



\maketitle The binary $\beta$-FeSe with the PbO structure is a key
member of the family of high-T$_c$ iron pnictide and chalcogenide
superconductors. Its structure comprises stacks of edge-sharing
FeSe$_4$ tetrahedra, similar to that of FeAs layers in the family of
the iron pnictide superconductors, but lacking charge-reservoir
layers. FeSe is thus the simplest Fe-based superconductor. The
undoped FeSe exhibits superconductivity with T$_c$=8 K\cite{MK Wu
11}. Upon applying pressure, T$_c$ dramatically rises to 37 K
\cite{11 p}. Recent studies on single layer FeSe thin film grown on
SrTiO$_3$ substrate by molecular beam epitaxy (MBE) method gave a
hint that the T$_c$ may even exceed 65 K \cite{QYWang,65-ARPES,65K,SYTan}.
It is widely expected that investigation of binary FeSe will provide
important clues to elucidate the superconducting mechanism of
iron-based superconductors.

Theoretical calculations indicate that $\beta$-FeSe has a two dimensional
electronic structure similar to that of the Fe-pnictides with
cylindrical electron sections at the zone corner and 
hole sections at the zone center\cite{Subedi,TX
theory,Moon}. Therefore, a similar spin density wave (SDW) order with a
collinear antiferromagnetic structure was predicted\cite{TX theory,Moon}.
However, no experimental evidence for the presence of such SDW order
was indicated in earlier experiments on polycrystalline
samples\cite{DJSingh}. Single crystal growth
of pure FeSe turns out to be difficult due to the very narrow phase
formation range\cite{SB Zhang,Cava2,MK3,Patel crystal 101,Petrovic 101}. The reported
plate-like single crystals grown by using ACl (A=Li, Na, K) as flux
usually contain secondary phase and quite often have a (101)
surface\cite{Petrovic 101}, making it very difficult to study the intrinsic in-plane
properties. As a result,
pure FeSe was far less studied than other Fe-based superconductors,
particularly by spectroscopic techniques. On the other hand, recent
studies indicate that the epitaxial growth of thin films on
SrTiO$_3$ substrate by MBE method could yield high quality (00l)
FeSe samples\cite{QYWang,65-ARPES,65K,SYTan}, making it ideal for studying in-plane
properties. Interestingly, a very recent
angle resolved photoemission spectroscopy (ARPES) measurement \cite{SYTan} on
such FeSe thin films indicated that, except for the single layered FeSe
film, the band structure at low temperature is very similar to those
observed for the parent compounds of FeAs-based systems, e.g.
BaFe$_2$As$_2$ and NaFeAs in the SDW state\cite{DL Feng 122split,DL
Feng 11SDW}, yielding evidence for
the formation of collinear antiferromagnetic structure. It was also
found that the SDW order temperature decreases with increasing the
thin film thickness, which was ascribed to the lattice relaxation
effect. The thick thin films (over 35 monolayers) resemble the bulk
samples but still have magnetic SDW ground state \cite{SYTan}.

\begin{figure}[b]
\includegraphics[width=2.7 in]{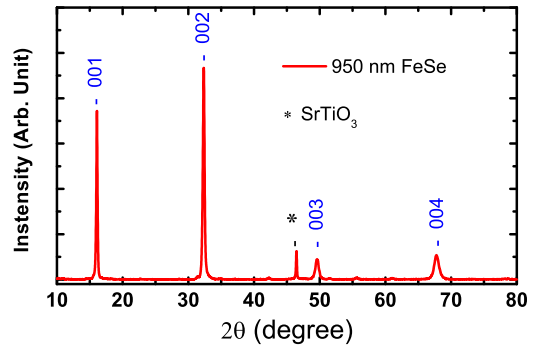}%
\vspace*{-0.20cm}%
\caption{\label{fig:R}(Color online) X-ray diffraction pattern of the 950 nm
FeSe thin film grown on (001) SrTiO$_3$ substrate.}
\end{figure}

Optical spectroscopy is a powerful bulk technique to investigate
charge dynamics and band structure of materials as it probes both
free carriers and interband excitations\cite{Dressel}. In
particular, it yields direct information about formation of energy
gaps. Infrared spectroscopy studies on the parent compounds
of Fe-pnictides (including systems in 122, 1111, 111) provide clear
evidence for the SDW gap formations in the ordered state\cite{WZ Hu
122,T Dong,WZ Hu 111}. It would be very interesting to
investigate the charge dynamic of binary $\beta$-FeSe to see if
similar gap opening effect exists by bulk probe technique. It should
be also very helpful to compare the charge dynamical properties of
FeSe with other Fe-pnictide and chalcogenide superconductors. In
this work, we present the first optical spectroscopy study on the
in-plane properties of $\beta$-FeSe thin film grown on SrTiO$_3$
substrate.

The thin film samples used in this study were obtained by the pulsed laser molecular beam epitaxy
on a (001) SrTiO$_3$ substrate. The advantage of this thin film growth technique is that it is
easy to grow relatively thick films. The laser energy, repetition, substrate 
temperature and pressure are 250 mJ/pulse, 10 Hz, 400$^0$C and 3$\times$10$^{-8}$ torr, 
respectively. The films had been deposited for 7 hours, resulting in a 
thickness of 950 nm. The epitaxy growth of the film is confirmed by the 
X-ray diffraction measurement, as shown in Fig. 1. Besides the
diffractions lines from the SrTiO$_3$ substrate, only FeSe (00l) diffraction peaks are present 
in the pattern. The pattern indicates that the film has an ab-plane surface. Note 
that the epitaxial growth of the film depends strongly on the
substrate. If the film is grown on the MgO substrate, the film would have a (101) crystal surface
\cite{MKfilm}.

\begin{figure}
\includegraphics[width=2.9 in]{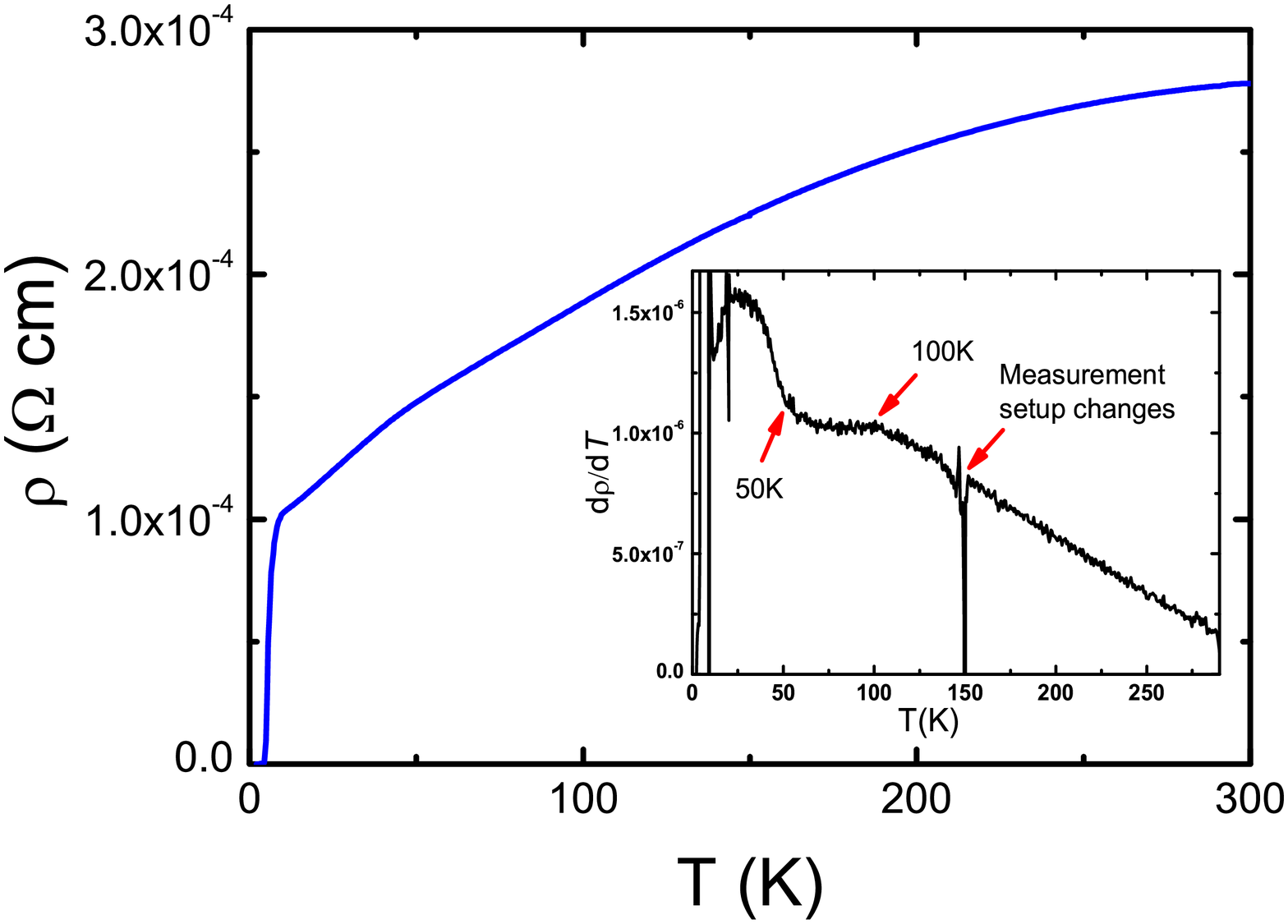}%
\vspace*{-0.20cm}%
\caption{\label{fig:R}(Color online) Temperature dependent resistivity
of FeSe film. The inset is the derivative of the resistivity as a function of temperature.}
\end{figure}

Figure 2 shows the in-plane resistivity measured in a Quantum Design
PPMS system. The resistivity shows a superconducting transition
temperature about 7 K, close to the bulk T$_c$. The inset shows the
temperature dependent derivative of the resistivity. A kink feature
could be observed near 100 K,
being similar to the report on polycrystalline samples \cite{YJSong}, which were
usually linked with the weak structural distortion. A further
increase of the slope is seen near 50 K, as evidenced more clearly
from the derivative plot.

The optical reflectance measurements were performed on a
combination of Bruker IFS 80v/s and 113v spectrometers in the frequency range
from 20 to 25000 cm$^{-1}$. An \textit{in situ} gold and aluminium
overcoating technique was used to get the reflectivity
R($\omega$). The real part of conductivity $\sigma_1(\omega)$ is
obtained by the Kramers-Kronig transformation of R($\omega$).

\begin{figure}
\includegraphics[width=2.9 in]{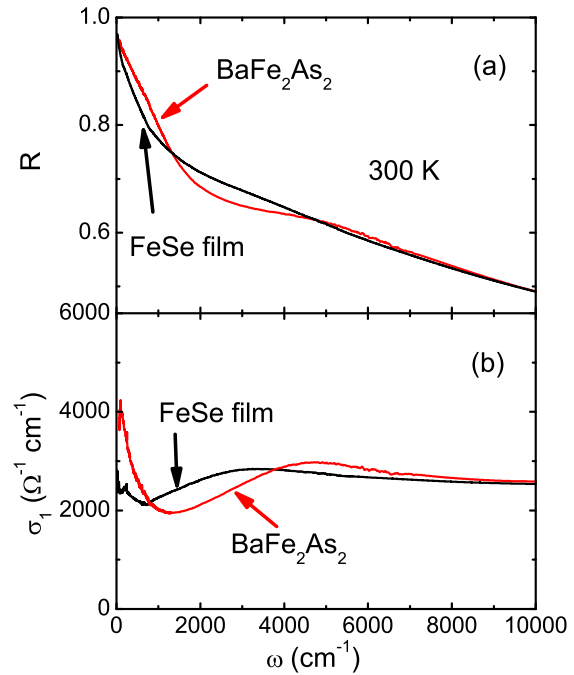}%
\vspace*{-0.20cm}%
\caption{\label{fig:R}(Color online) The room temperature optical reflectance R($\omega$) (a) and
conductivity $\sigma_1(\omega)$ (b) over broad frequencies. Data from the BaFe$_2$As$_2$ \cite{WZ Hu 122} are also
shown for comparison.}
\end{figure}

Figure 3 shows the room temperature optical reflectance R($\omega$) (upper panel) and
conductivity $\sigma_1(\omega)$ (lower panel). No leakage from the SrTiO$_3$ substrate is 
detected. The overall spectral lineshapes are similar
to iorn pnictides, e.g. BaFe$_2$As$_2$. As a comparison, we
have included the room-temperature reflectance data of 
BaFe$_2$As$_2$ in the figure \cite{WZ Hu 122}.
The reflectance drops quickly with frequency at
low-$\omega$ region, then merges into the high values of a background
contributed mostly from the incoherent carriers and interband transitions.
However, a clear difference between FeSe and BaFe$_2$As$_2$ could be identified:
the reflectance of FeSe at low frequency range (below 1200 \cm) is apparently lower than that of
BaFe$_2$As$_2$, but reversed at higher frequencies. This leads to smaller
free carrier spectral weight but enhanced spectral weight in
the frequency range between 1000 $\sim$ 4000 \cm
in $\sigma_1(\omega)$.

The temperature dependent optical reflectance R($\omega$) and
conductivity $\sigma_1(\omega)$ are shown in Fig. 4. The upper
panels ((a) and (b)) show the spectra over broad frequencies (up to
8000 \cm, $\sim$ 1 eV), the lower panels ((c) and (d)) are the
expanded plots at low frequencies. From the lower panels, we can see
a phonon mode at 240 \cm at room temperature, which shifts slightly
to higher frequency at low temperature (248 \cm at 8 K) and also
becomes more visible. The phonon mode is commonly seen in the
in-plane infrared measurement on FeAs(Se)-based single crystal
samples and ascribed to the in-plane displacements of Fe-As(Se)
atoms \cite{phonon 122}. The mode appears at higher frequency than that
observed at 187 \cm for Fe$_{1.03}$Te and 204 \cm for
FeTe$_{0.55}$Se$_{0.45}$ samples \cite{homes 11}. This is due to the reduced mass of
Se atom as compared with Te atom. For FeAs-based compounds, the mode
appears at slightly higher energy scale, e.g. 253 \cm for
BaFe$_2$As$_2$ at 300 K \cite{WZ Hu 122,phonon 122}. The clear
observation of this in-plane phonon mode is an indication of good
quality of the sample.

\begin{figure}
\includegraphics[clip,width=1.65in]{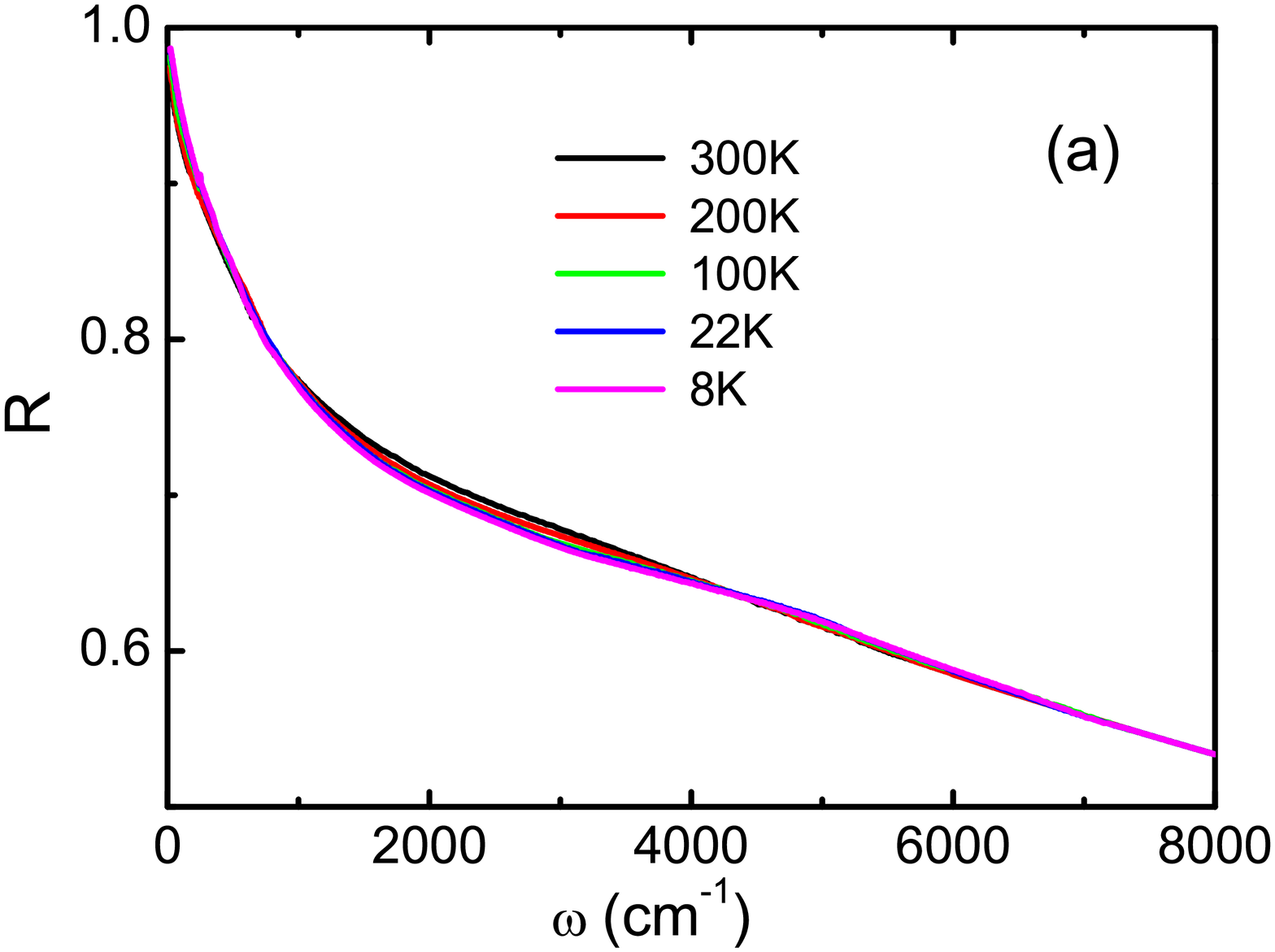}
\includegraphics[clip,width=1.7in]{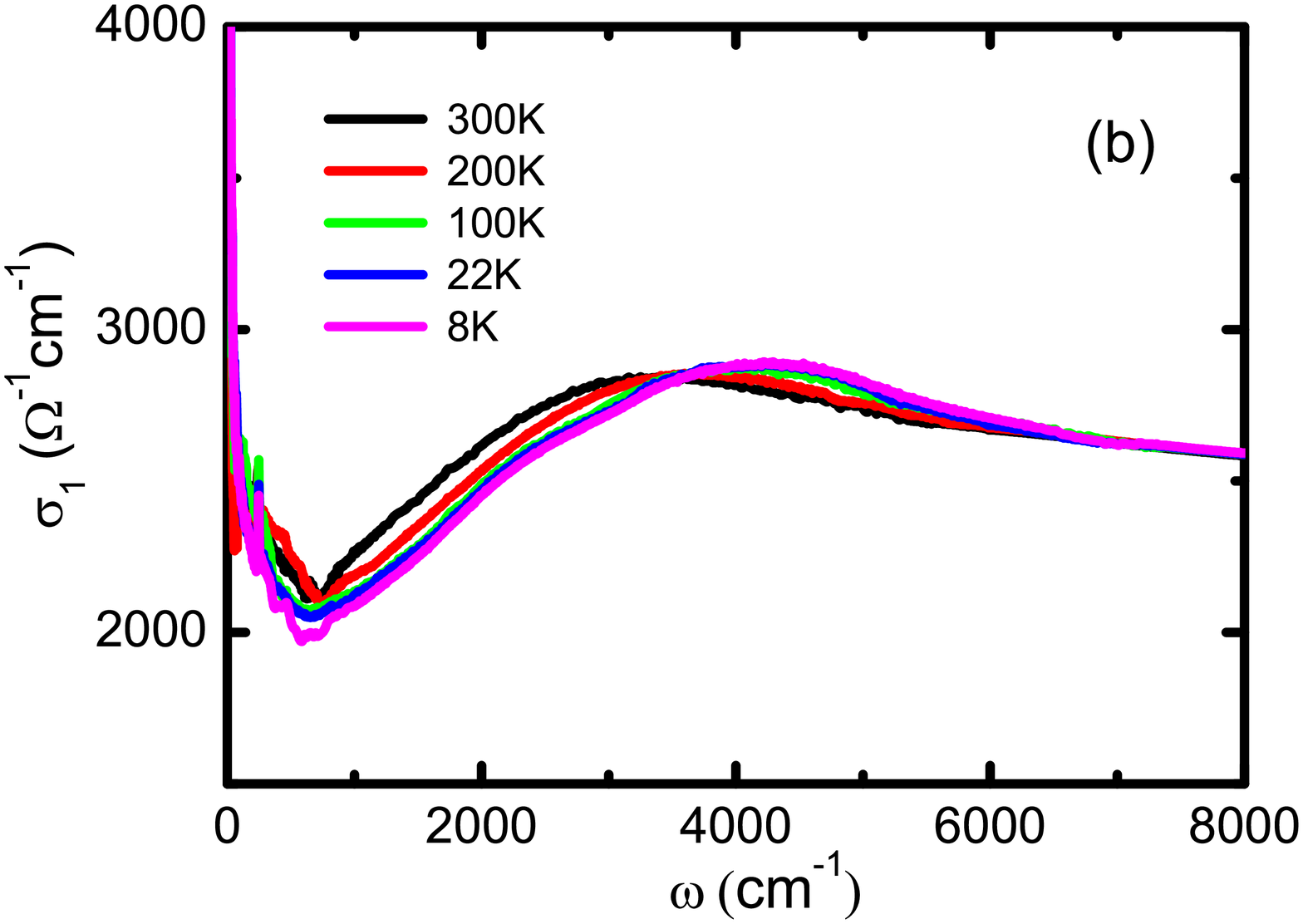}
\includegraphics[clip,width=1.65in]{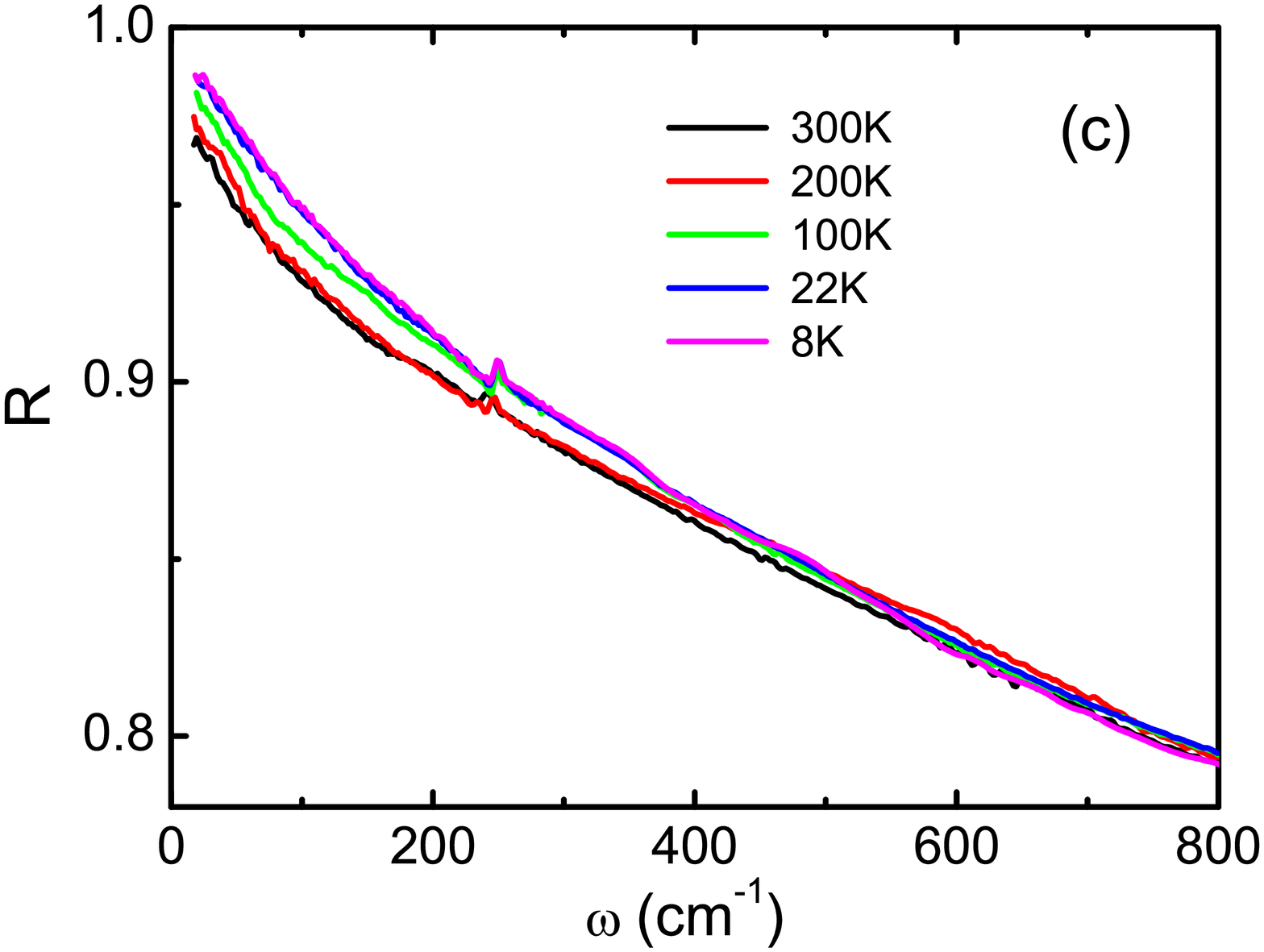}
\includegraphics[clip,width=1.7in]{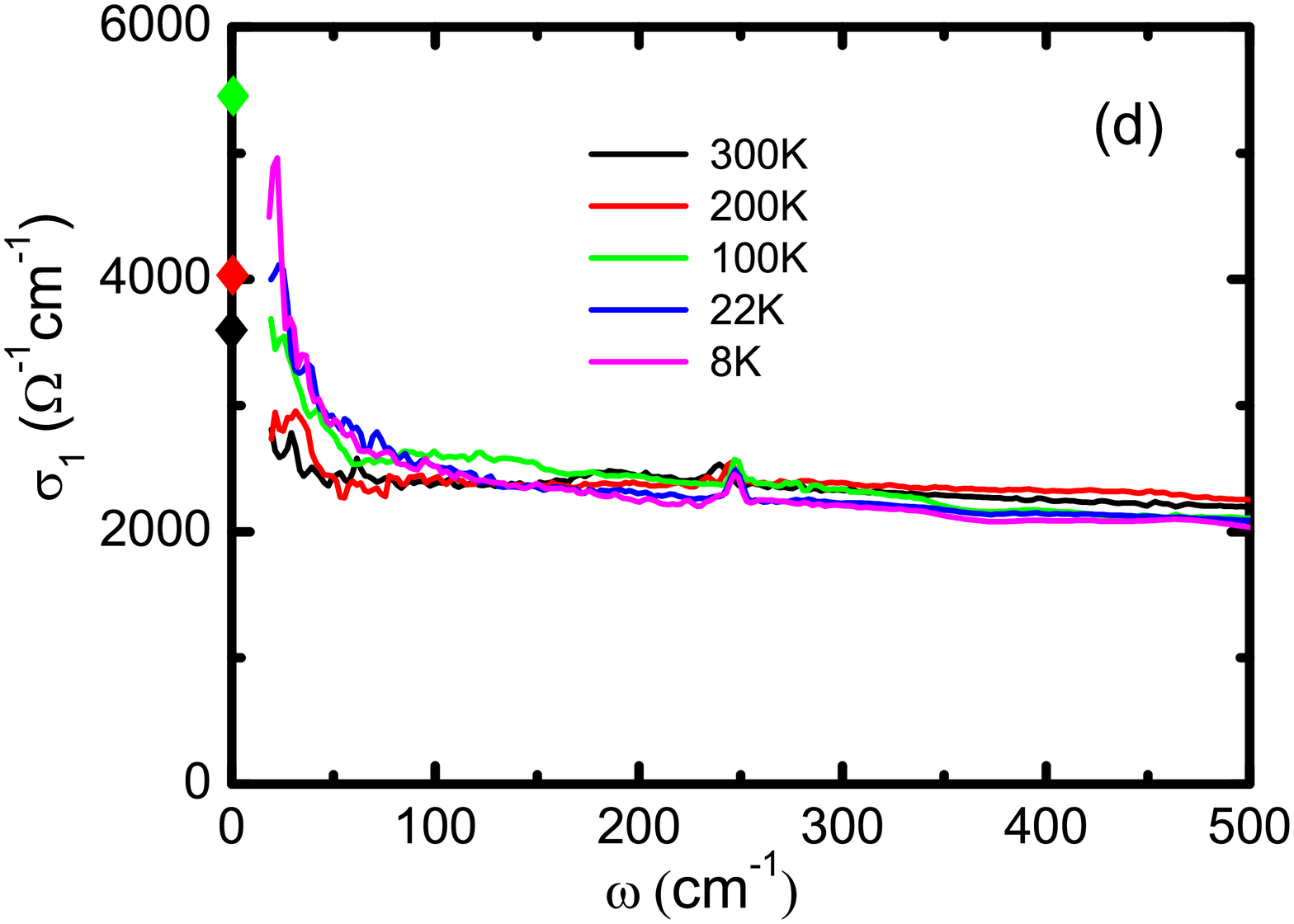}
\caption{The reflectance R($\omega$) and conductivity spectra
$\sigma_1(\omega)$ of FeSe film at different temperatures. Upper panels: The
spectra over a broad frequencies up to 8000 \cm.
Lower panels: The expanded plot of R($\omega$) and $\sigma_1(\omega)$ in the low
frequency region. The dc conductivity values at three temperatures
300, 200 and 100 K were added.}
\end{figure}

It is tempting to estimate the Drude weight or plasma frequency of
the FeSe sample and compare it with other iron
pnictides/chalcogenides. The $\sigma_1(\omega)$ spectrum shows very
weak frequency-dependent behavior below roughly 700$\sim$800 \cm.
However, a sharp and narrow component develops below 100$\sim$150
\cm (see Fig. 3 (d)). The spectra are more similar to FeTe or
FeTe$_{0.55}$Se$_{0.45}$ single crystals \cite{GFChen,homes 11} at low temperature than
BaFe$_2$As$_2$ or other Fe-pnictides \cite{WZ Hu 122,phonon 122,T Dong,WZ Hu 111}. 
If we consider only the sharp
and narrow component below 100$\sim$150 \cm as the Drude component,
the slowly decreasing $\sigma_1(\omega)$ spectrum above it as
incoherent, we may underestimate the plasma frequency. So we assume
two components contribute to the Drude spectral weight: an extremely
narrow one below 150 \cm and a broad one contribute to the rather
slowly decreasing conductivity between 150 and 800 \cm. Then, we can
analyze the $\sigma_1(\omega)$ data using a Drude-Lorentz model in a
way similar to what we did for 122-type crystals, \cite{WZ Hu 122}
\begin{equation}
\epsilon(\omega)=\epsilon_\infty-\sum_{i}{{\omega_{p,i}^2}\over{\omega_i^2+i\omega/\tau_i}}+\sum_{j}{{\Omega_j^2}\over{\omega_j^2-\omega^2-i\omega/\tau_j}}.
\label{chik}
\end{equation}
where $\epsilon_\infty$ is the dielectric constant at
high energy, and the middle and last terms are the Drude and Lorentz components,
respectively.

\begin{figure}
\includegraphics[clip,width=1.65in]{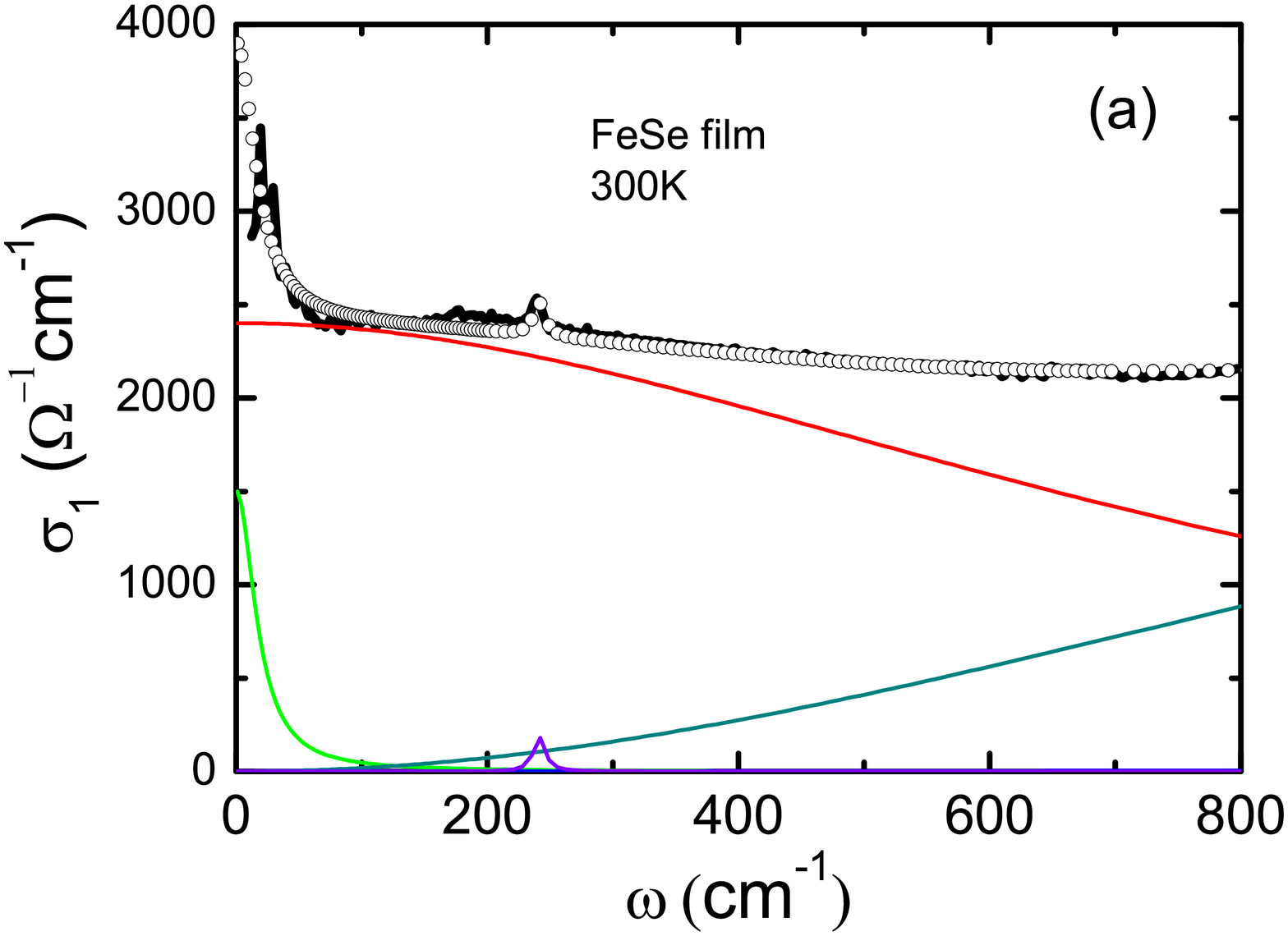}
\includegraphics[clip,width=1.65in]{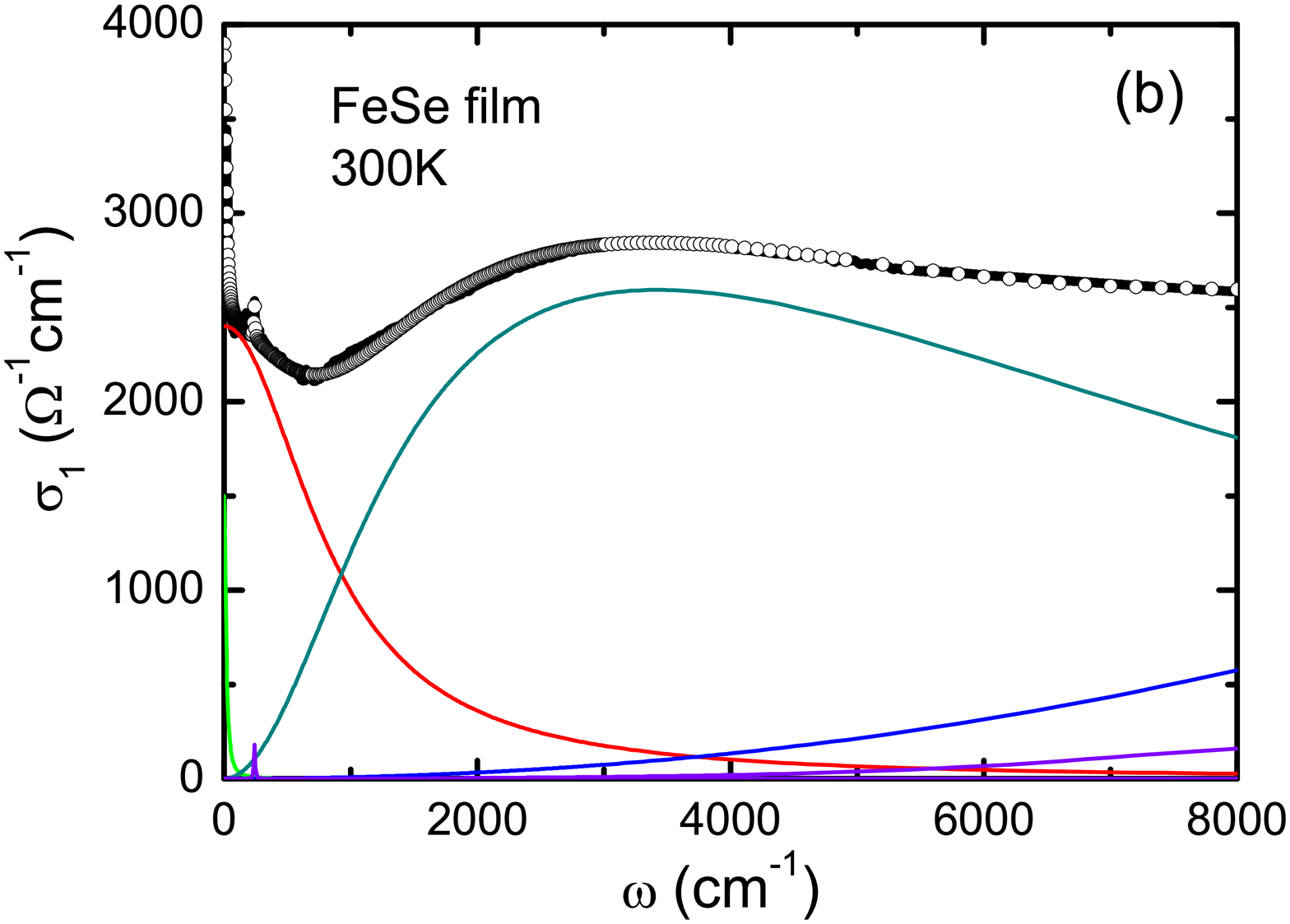}
\caption{The experimental data of $\sigma_1(\omega)$ at 300 K FeSe film and the Drude-Lorentz fit results.
(a) The data and fitting curve at low frequency range, and (b) the data and fitting curves at
broad frequency range. }
\end{figure}

We find that the optical conductivity data could be reasonably
reproduced by this approach. Figure 4 shows the $\sigma_1(\omega)$
data at 300 K together with the components in the Drude-Lorentz
analysis. Panel (a) shows the spectral data at low frequencies,
while panel (b) shows the data over broad frequencies up to 8000
\cm. The parameters of two Drude components are: $\omega_{p1}$=1300
\cm, $1/\tau_1$=20 \cm, and $\omega_{p2}$=10900 \cm, $1/\tau_1$=840
\cm at 300 K. At 8 K, those parameters become $\omega_{p1}$=2700
\cm, $1/\tau_1$=16 \cm, and $\omega_{p2}$=10600 \cm, $1/\tau_1$=830
\cm. Then, the overall plasma frequency $\omega_p$ could be
calculated as
$\omega_p$=($\omega_{p1}^2$+$\omega_{p2}^2$)$^{1/2}$$\approx$11000
\cm (1.36 eV) for different temperatures. This value is somewhat
smaller than that in BaFe$_2$As$_2$ (about 1.6 eV)\cite{WZ Hu 122}.
In density functional calculations (DFT), the plasma frequency of
FeSe is slightly higher than 3 eV \cite{ZPYin}. In comparison with this value, we
find the band renormalization factor
$\omega_{p,DFT}^2$/$\omega_{p,exp}^2$$\sim$ 5, which is larger than
BaFe$_2$As$_2$. On the other hand, the value is close to that
obtained by a combination of density functional theory and dynamical
mean field theory where electron correlation effect has been
properly taken into account\cite{ZPYin,Ai theoretical}. The agreement
between experimentally obtained values and those from such
theoretical calculations for Fe pnictides was already well
documented \cite{ZPYin}.

The plasma frequency could be alternatively estimated by summarizing the low-$\omega$ spectral weight,
$\omega_p^2$=8$\int^{\omega_c}_0\sigma_1(\omega)d\omega$. The integration
up to $\omega_c$ should cover all the spectrum contributed by the free
carriers but still below the inter-band transition. Usually, one takes $\omega_c$
at the frequency where the $\sigma_1(\omega)$ shows a minimum, then we
expect that there is a balance between the Drude component tail
and the onset part of the interband transition. Taking $\omega_c$$\approx$800 \cm,
we get $\omega_p\approx$8400 \cm. This value is smaller than the value obtained from the above two
Drude component analysis. This is because the tail of the broad Drude component is not
balanced by the onset part of the Lorentz component, as can be seen in Fig. 4 (b).
Therefore, taking $\omega_c$ at the energy that $\sigma_1(\omega)$
shows a minimum in the present case still leads to an underestimation of plasma frequency.

Although presence of multiple Drude components appears to be the generic property of iron
pnictides/chacolgenides, the development of narrow and sharp Drude component at such low
frequencies were not seen in iron pnictide systems \cite{WZ Hu 122,phonon 122,T Dong,WZ Hu 111}. 
It suggests that in certain peculiar band
of FeSe the quasiparticles experience extremely small scattering at low temperature.

Another prominent feature in optical conductivity (Fig. 3 (b)) is that
the low frequency spectral weight is transferred to the high energy
region above 4000 \cm with decreasing temperature. Such spectral
weight transfer feature was also observed in iron pnictides systems \cite{wang
Hund,Schafgans}, and was referred to as a "high-energy psuedogap" structure \cite{wang
Hund}. But the relevant energy scale is smaller for FeSe. For BaFe$_2$As$_2$, the
spectral weight was transferred to region above 5000 \cm. 
The temperature induced spectral weight transfer was ascribed
to the electron correlation effect, in particular, to the Hund's
coupling effect between itinerant Fe 3d electrons and localized Fe
3d electrons in different orbitals \cite{wang
Hund,Schafgans}. So, the spectral weight transfer represents the redistribution 
of the spectral weight between different 3d bands. It is interesting to note that
FeSe as well as other iron calcogenide compounds have stronger
electron correlation effect, leading to larger local moment of
2$\mu_B$ \cite{Gretarsson} and higher band renormalization factor. However, the energy
scale of temperature induced spectral weight transfer in optical
conductivity appears to be smaller. This puzzling phenomenon reflects the 
complex interplay effect from correlations/Hund's couplings and kinetic energy of 
electrons in FeSe(As) systems and needs to be further explored.

As we mentioned above, a recent ARPES measurement on (001) FeSe thin film \cite{SYTan} indicated
signature of SDW order, in a way very similar to the parent compound of Fe pnictides, e.g. BaFe$_2$As$_2$.
It would be highly desirable to examine this possibility by optical measurement. From
Fig. 4 (c), we find that the reflectance R($\omega$) at low frequency
shows a monotonous increase with decreasing temperature without showing any
specific suppression arising from energy gap formation. At two lowest measurement
temperatures 8 K and 22 K, the reflectance seems to develop a stronger upward curvature below
the phonon mode frequency, being consistent with the even faster decrease of the dc
resistivity below 50 K. Nevertheless, no any energy gap signature could be identified from the
reflectance spectra. As a result, in the $\sigma_1(\omega)$ spectra, we observe a further
narrowing of Drude component at low temperature. Since the infrared measurement is a bulk
detection technique and infrared experiments on all parent compounds of iron pnictides
(including 122, 111, 1111 systems) revealed clearly formation of energy gaps in the SDW state or
even below the structural distortion, our measurement indicates that the superconducting
bulk FeSe film is not in the SDW state.

It is worth remarking that an earlier pump-probe experiment on (101)
FeSe film also indicated a gapped relaxation of excited
quasiparticles below roughly 100 K \cite{MKfilm}. However, the
extraction of an energy gap from this technique is indirect. The
quasiparticles were excited by the pump laser (3.1 eV) to states
far above the Fermi level, then the transient reflection was probed at the energy of 1.55 eV
on those excited quasiparticles in the relaxation process towards
their equilibrium state. The probe is different from the optical
measurement in the equilibrium state. Anyhow, our measurement can
not rule out that such films or those grown on SrTiO$_3$ substrate
with thickness of a few tens of monolayers have SDW orders at low
temperature.

In summary, we have investigated the ab-plane optical properties of
FeSe films grown on SrTiO$_3$ substrate and compared with Fe
pnictides, e.g. BaFe$_2$As$_2$. The low frequency conductivity
spectrum consists of two components: a broad one which takes up most
of the spectral weight and a narrow one roughly below 100-150 \cm.
The narrow Drude component locates at so low frequencies that no
such behavior was observed in iron pnictide systems. The study
revealed a smaller plasma frequency and enhanced renormalization
effect in FeSe. Furthermore, different from the recent ARPES
measurement which revealed a SDW order at low temperature for
relatively thick thin films grown on SrTiO$_3$ substrate, the
present optical measurement does not yield any sign of the energy
gap formation in the thick film.

\begin{acknowledgments}
This work was supported by the National Science Foundation of China,
and the 973 project of the Ministry of Science and Technology of
China (2011CB921701,2012CB821403).
\end{acknowledgments}


\end{document}